\documentclass[aps,prd,nofootinbib]{revtex4}
\usepackage{graphicx}

\begin{document}
\title{Reply to ``Cuts and penalties: comment on `The clustering of ultra-high 
energy cosmic rays and their sources' ''} 
\author{N.W. Evans $^1$, F. Ferrer $^2$
and S. Sarkar $^2$}
\medskip
\affiliation{$^1$ Institute of Astronomy, University of Cambridge,
Madingley Road, Cambridge CB3 OHA, UK \\ 
$^2$ Theoretical Physics, University of Oxford, 1 Keble Road,
Oxford OX1 3NP, UK} 

\begin{abstract}
We reiterate that there is no evidence that BL Lacs are sources of
ultrahigh energy cosmic rays.
\end{abstract}

\pacs{98.70.Sa, 95.35.+d}
\maketitle

Tinyakov and Tkachev (TT) \cite{Tinyakov:2001nr} have claimed that
{\em ``BL Lacertae are sources of the observed ultra-high energy
cosmic rays''} (UHECRs). They considered a set of 39 UHECRs with $E >
4.8 \times 10^{19}$~eV observed by AGASA and 26 UHECRs with $E > 2.4
\times 10^{19}$~eV observed by Yakutsk, and compared their arrival
directions with the positions of 22 BL Lacs selected by redshift ($z >
0.1$ or unknown), apparent magnitude ($m < 18$) and 6 cm radio flux
($F_6 > 0.17$~Jy). Eight UHECRs were found to be within $2.5^\circ$ of
5 BL Lacs, the chance probability of which was estimated to be
$6\times10^{-5}$ including all penalties for the arbitrary cuts made
\cite{Tinyakov:2001nr}. We have shown \cite{Evans:2002ry} that the
significance of the coincidences has been greatly exaggerated. In the
preceding {\em Comment} \cite{Tinyakov:2003} TT assert that our
criticism is incorrect. We argue below that this is not the case and
provide further evidence in support of our position.

Our first criticism was that TT did not take into account the (energy
dependent) angular resolution of the experiments. Although the
positions of the BL Lacs are known to arcsecond accuracy, the arrival
directions of UHECRs in air shower arrays cannot be reconstructed to
better than a few degrees. In particular for simulated events in
AGASA, 68\% have a reconstructed arrival direction within $1.8^\circ$
of the true direction and 90\% within $3^\circ$; the corresponding
angles for all events above $10^{19}$~eV are $2.8^\circ$ and
$4.6^\circ$ \cite{Takeda:1999sg}. TT require, without providing
specific justification, that the UHECR arrival direction be within
$2.5^0$ of a BL Lac in order to be considered a coincidence. This may
appear to be a reasonable approximation for the AGASA data. When it
comes to the Yakutsk data however, the angular resolution is far worse
for the lower energy events considered, in particular it exceeds $4^0$
for $E < 4 \times 10^{19}$~eV
\cite{yakcat,Uchihori:1999gu}. Nevertheless the most significant
correlation listed by TT is that of a `triplet' of UHECRs in the
Yakutsk data having energies of (3.4, 2.8, 2.5) $\times10^{19}$~eV
whose nominal arrival directions are within $2.5^0$ of a BL Lac (1ES
0806+524). In their {\em Comment} \cite{Tinyakov:2003}, TT assert:
{\em ``By itself, worse angular resolution does not imply that
correlations with sources must be absent in Yakutsk set: even though
the angular resolution is worse, the density of UHECR events around
actual sources is larger as compared to a random set, and one has
excess in counts even at small angles''}. If this were indeed the
case, then one would reasonably expect UHECRs observed by other
experiments (with better angular resolution) to be (even better)
aligned with the BL Lacs in question. In fact there are {\em no} such
coincidences with any of the 39 AGASA events they considered!
Therefore we reassert that there is no justification for ascribing any
significance to coincidences between Yakutsk events and BL Lacs within
$2.5^0$.

\begin{figure}[hbt]
\begin{center}
\begin{tabular}{c@{\hspace{2cm}}c}
\includegraphics[width=7.5cm,height=5.5cm]{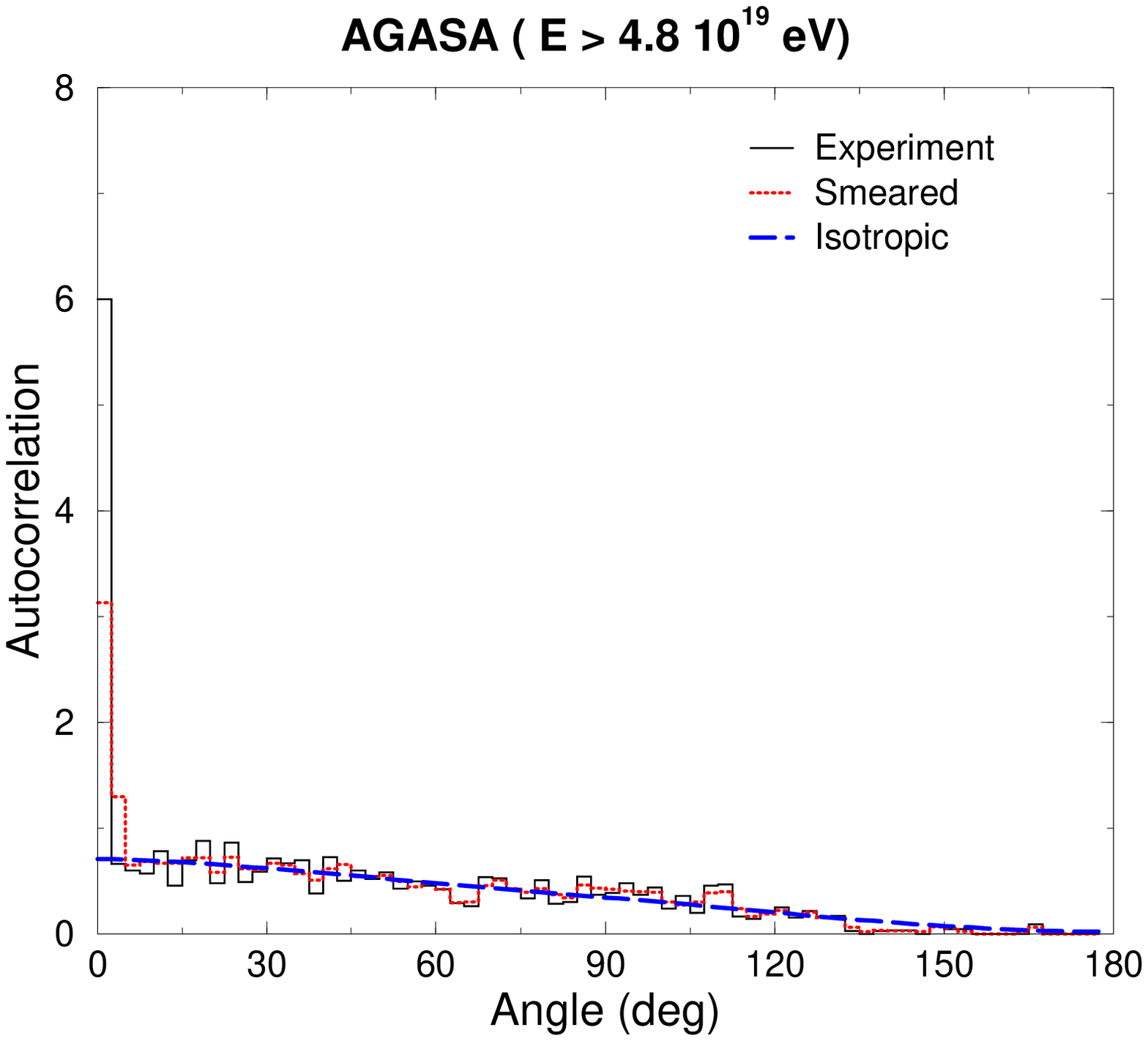}& 
\includegraphics[width=7.5cm,height=5.5cm]{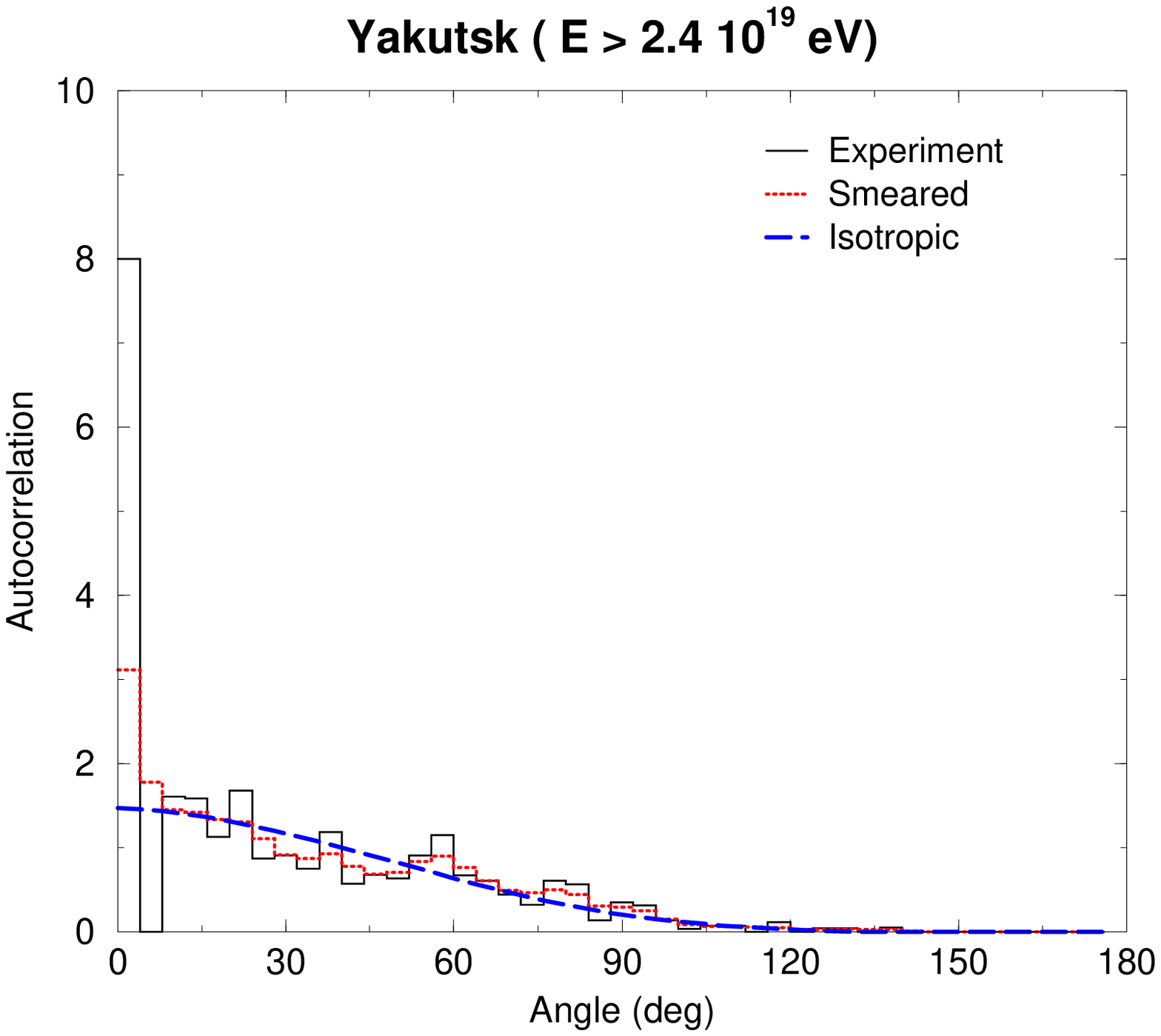}
\end{tabular}
\end{center}
\caption{Autocorrelation for AGASA and Yakutsk}
\label{autocorr}
\end{figure}

To demonstrate this quantitatively we have calculated the
autocorrelation functions of the selected AGASA and Yakutsk events
\cite{Tinyakov:2001ic}, as well as their cross-correlation with the 22
selected BL Lacs \cite{Tinyakov:2001nr}, taking the angular resolution
of the experiments into account. For each observed UHECR, a new
arrival direction is generated from the distribution defined by the
quoted experimental angular resolution at that energy, as has been
done e.g. for BATSE data \cite{batse}. We generate $10^6$ such data
sets, for comparison with the data sets generated from an isotropic
distribution. As seen in Fig.~\ref{autocorr}, this has a dramatic
effect on the significance of the claimed clustering. We find the
chance probability for an isotropic distribution to yield as many
events (with $E > 4.8 \times 10^{19}$ eV) as was observed by AGASA in
the first ($2.5^0$) angular bin to be $1.8\times10^{-4}$. Similarly
the chance probability for an isotropic distribution to yield as many
events (with $E > 2.4 \times 10^{19}$ eV) as was observed by Yakutsk
in the first ($4^0$) angular bin is $6.5 \times 10^{-4}$. Both these
numbers agree with TT's estimates in Table 1 of
ref.\cite{Tinyakov:2001ic}, allowing for their `penalty factor' of
$\sim3$. However when we take the angular smearing into account, these
chance probabilities increase to 3.5\% for AGASA and 18\% for
Yakutsk. Thus there is little basis for the claim that {\em
``Correlation function of ultrahigh energy cosmic rays favours point
sources''} \cite{Tinyakov:2001ic}. The significance of the clustering
in the AGASA data has also been questioned recently by other authors
\cite{Finley:2003ur}; however they did not take the limited angular
resolution of AGASA into account.

\begin{figure}[hbt]
\begin{center}
\begin{tabular}{c@{\hspace{2cm}}c}
\includegraphics[width=7.5cm,height=5.5cm]{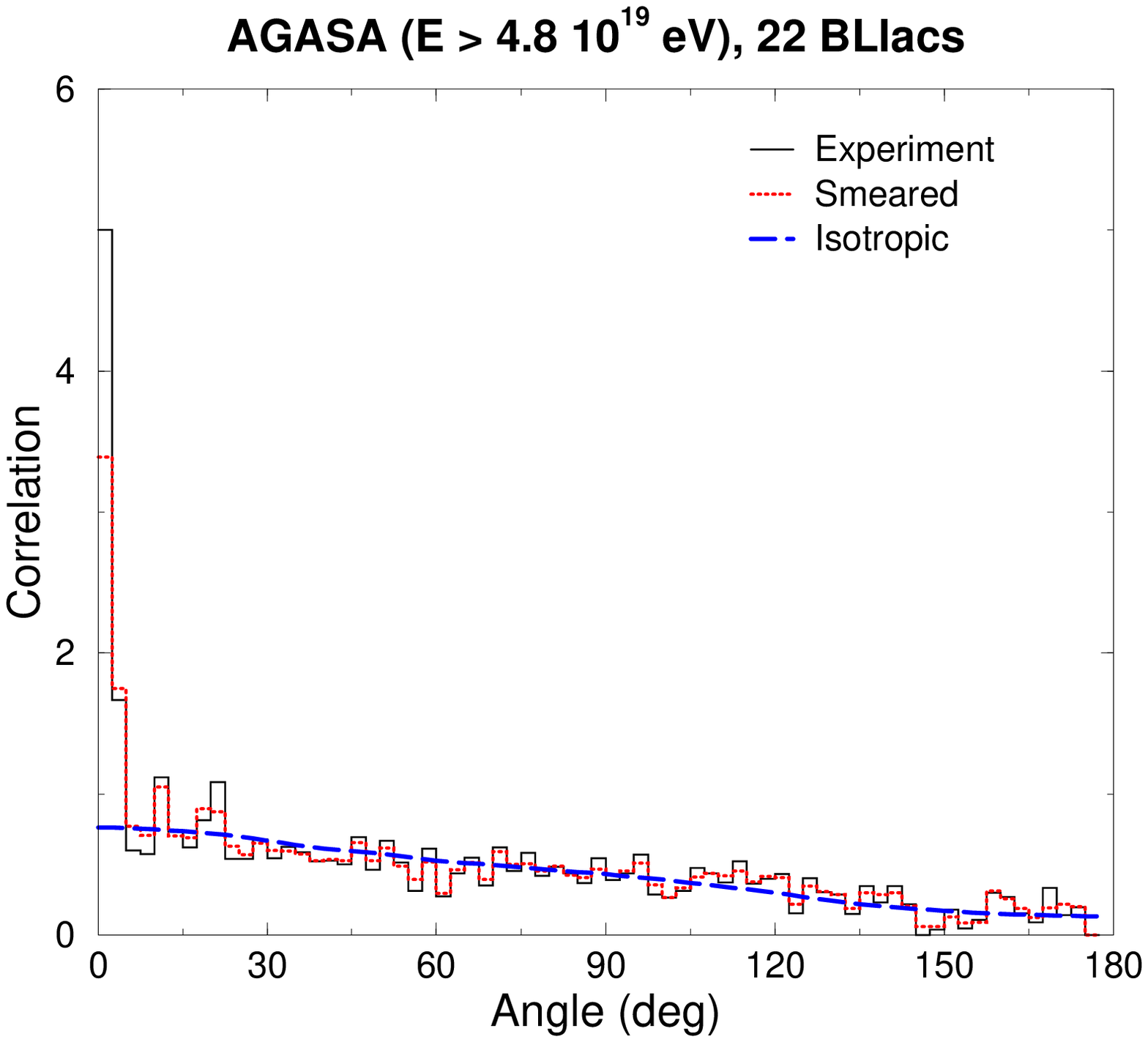}&
\includegraphics[width=7.5cm,height=5.5cm]{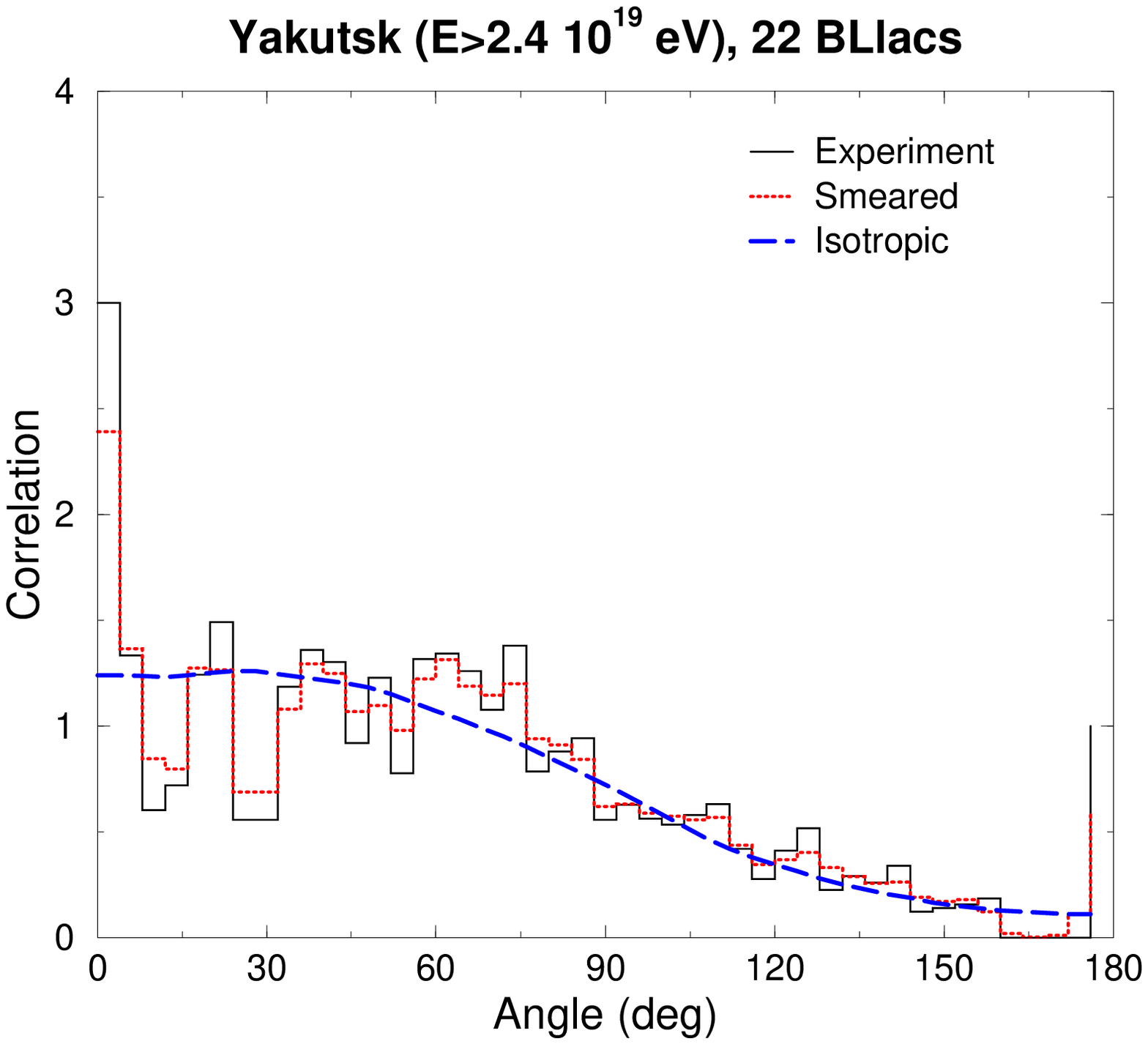}
\end{tabular}
\end{center}
\caption{Cross-correlation with selected BL Lacs for AGASA and Yakutsk}
\label{blcorr}
\end{figure}

Concerning the cross-correlation with the 22 BL Lacs selected by TT ,
the probability for an isotropic distribution of UHECRs to yield as
many coincidences between the AGASA events and these BL Lacs as is
actually observed, is only $1.5 \times 10^{-3}$, but this chance
probability increases to 4\% when the angular smearing is taken into
account. For the Yakutsk data, the chance probability is
$8\times10^{-2}$ without the angular smearing, but as high as 38\%
when this is included. Thus as shown in Fig.~\ref{blcorr}, there is
{\em no} justification for TT's inclusion of the Yakutsk data; they do
so simply because when the AGASA and Yakutsk datasets are combined,
new clusters appear combining events from both datasets, thus
artificially enhancing the significance of the coincidences.

Our second criticism was directed at TT's assumption that {\em ``
\ldots the energies of the events are not important for correlations
at small angles \ldots ''} \cite{Tinyakov:2001nr}. We demonstrated
\cite{Evans:2002ry} that by lowering the energy cut on the AGASA data
from $4.6 \times 10^{19}$~eV to $4 \times 10^{19}$~eV, the
significance of the coincidences in fact {\em decreases} by a factor of 5.

In closing we would like to draw attention to other recent papers
which have a bearing on this issue. Using an independent sample of 33
UHECRs observed by Volcano Ranch and Haverah Park, {\em no}
coincidences are found between their arrival directions and the 22 BL
Lacs selected by TT \cite{Tinyakov:2001nr}; the probability that this
null result arises as a fluctuation from the strongly correlated case
is less than 5\% \cite{Torres:2003ee}. Secondly an independent
analysis of the AGASA events finds {\em no} statistically significant
correlations with BL Lacs \cite{Burgett:2003gx}.


\end{document}